\documentclass[showpacs,preprintnumbers,two column]{revtex4}
\usepackage{amsmath}
\usepackage{graphicx}
\usepackage{dcolumn}
\usepackage{bm}

\begin{document}

\title{Generalized rotating-wave approximation to the two-qubit and cavity coupling
system}
\author{Yu-Yu Zhang$^{1}$, Qing-Hu Chen$^{2}$}

\address{
$^{1}$Department of Physics, Chongqing University, Chongqing
400044, P. R.  China\\
$^{2}$Department of Physics, Zhejiang University, Hangzhou 310027,
P. R. China
}\date{\today}

\begin{abstract}
The generalized rotating-wave approximation (GRWA) is presented for the
two-qubit and cavity coipling system . The analytical expressions in the
zeroth order approximation recover the previous adiabatic ones. The
counterrotating-wave terms can be eliminated by performing the first order
corrections. An effective solvable Hamiltonian with the same form as the
ordinary RWA one are then obtained, giving a significantly accurate
eigenvalues and eigenstates. Energy levels in the present GRWA are in
accordant with the numerical exact diagonalization ones in the a wide range
of coupling strength. The atomic population inversion in the GRWA is in
quantitative agreement with the numerical results for different detunings in
the ultrastrong coupling regime.
\end{abstract}

\maketitle

\section{introduction}

Recent experimental progress related to qubit-oscillator systems using
superconducting qubit circuits has made it possible to achieve the so-called
ultrastrong-coupling regime, where the coupling strength between a single
qubit and a single oscillator is comparable to the bare frequencies of the
two constituents~\cite{Niemczyk,pfd,fedorov,devoret,you}. In this regime,
the ubiquitous rotating-wave approximation (RWA)~\cite{jaynes} is expected
to break down, leading to a mass of unexplored physics and giving rise to
fascinating quantum phenomena, such as the asymmetry of vacuum Rabi
splitting~\cite{cao,zhu}, collapse and revival dynamics~\cite{casanova,Braak}%
, a Bloch-Siegert shift~\cite{pfd}, super-radiance transition~\cite
{Ashhab,lambert,zhang}, and radiation processes based on virtual photons~%
\cite{ciuti,law,stassi}. It is highly desirable to understand the behavior
of the qubit-oscillator in the whole coupling regime.

Since the Hamiltonian of a qubit-oscillator system contains counter
rotating-wave terms, the total bosonic number is not conserved, it is very
changeling to obtain the analytical solutions in the ultrastrong-coupling
regime. There is much on-going interest in this field. The Rabi model ~\cite
{rabi} describing a single qubit interacting with a quantum harmonic
oscillator has been studied extensively beyond RWA with various analytical
methods~\cite{irish,braak1,liu} in the recent years. Two or more qubits
coupled to a common harmonic oscillator in the ultrastrong-coupling regime
has more potential applications in quantum information processing than that
in the single-qubit Rabi model, such as the implement quantum-information
protocols with the oscillator transferring information coherently between
qubits~\cite{leek}, and quantum entanglement of multiqubit properties, and
superradiance phase transition in the Dicke model describing two-level atoms
ensemble in a cavity~\cite{lambert,chen1}. We investigate the Tavis-Cummings
model beyond the RWA, in which a quantum harmonic oscillator interacts with
two identical qubits symmetrically. One of the motivations lies in the
absence of extensively study of the two- and more-qubit in the
ultrastrong-coupling regime. Recently, an adiabatic approximation functions
well when the qubit frequency is much smaller than the oscillator frequency~%
\cite{agarwal}, and the variational treatment~\cite{lee} reasonably captures
the properties of the ground state in the Tavis-Cummings model.

We focus here on the analytic energy spectrum and eigenstates of the
Tavis-Cummings model with two identical qubits beyond the RWA in the
ultrastring-coupling regime by the generalized rotating-wave approximation
(GRWA). By mapping the Tavis-Cummings with counterrotating-wave interactions
into a solvable Hamiltonian with the same form as the ordinary RWA term, we
show that all eigenvalues and eigenstates can be approximated determined by
the analytical expression based on our method, which agrees well with the
exactly numerical simulation in the ultrastrong coupling regime under
different detunings. We recovers the same results with zero order
approximation as that in Ref.~\cite{agarwal}, and make great improvement of
energy spectrum by the first order corrections. The two-qubit population
dynamics is calculated to justify the validity of the eigenstates within a
wide range of parameters.

\section{Hamiltonian and zero order approximation}

The Hamiltonian of the Tavis-Cummings model, where two identical qubits
couple to a harmonic oscillator with the counter rotating-wave interaction,
is $\left( \hbar =1\right) $
\begin{eqnarray}  \label{crtham}
H =\Delta J_{x}+\omega a^{\dagger }a+g(a^{\dagger }+a)J_{z},
\end{eqnarray}
where $a$ and $a^{\dagger}$ are, respectively, the annihilation and creation
operators of the harmonic oscillator with frequency $\omega$. The collective
spin-$1$ angular momentum operators $J_{z}=\frac{1}{2}(\sigma_{z}^{1}+%
\sigma_{z}^{2})$ and $J_{x}=\frac{1}{2}(\sigma_{x}^{1}+\sigma_{x}^{2})$.
Physically, the spin-$1$ system can be formed by the two identical qubits
in their triplet space. $\Delta$ is the atomic transition frequency, and $g$
denotes the collective qubit-oscillator coupling strength.

To begin with, a brief review of the standard RWA is given in order to
establish the arguments used in deriving the generalized approximation. The
first step is to rewrite Eq.(~\ref{crtham}) in the form
\begin{eqnarray}  \label{crtham1}
H =-\Delta J_{z}+\omega a^{\dagger }a+\frac{g}{2}(a^{%
\dagger}+a)(J_{+}+J_{-}),
\end{eqnarray}
where $J_{\pm}$ are the collective atomic raising and lowering operators of
a spin-{1} system. In the basis $|j_{z}=1, n-1\rangle $, $|j_{z}=0,n\rangle $
and $|j_{z}=-1, n+1\rangle $, which is the eigenstates of the noninteracting
Hamiltonian $-\Delta J_{z}+\omega a^{\dagger }a$, the interaction term $%
a^{\dagger}J_{-}+aJ_{+}$ couples the states $|j_{z}=1, n-1\rangle$ with $%
|j_{z}=0,n\rangle $, and $|j_{z}=0,n\rangle $ with $|j_{z}=-1, n+1\rangle $,
where energy is conserved. On the other hand, the counter rotating-wave
terms $a^{\dagger}J_{+}+aJ_{-}$ couples the off-resonant states, such as $%
|j_{z}=0, n\rangle$ with $|j_{z}=1,n+1\rangle $ and $|j_{z}=-1,n-1\rangle $,
where energy is non-conserved. To eliminate the counter rotating-wave terms,
the RWA Hamiltonian $H_{RWA}=-\Delta J_{z}+\omega a^{\dagger }a+\frac{g}{2}%
(a^{\dagger }J_{-}+aJ_{+})$ can be written a matrix form
\begin{eqnarray}  \label{RWA}
H_{RWA} =\left(
\begin{array}{ccc}
\omega (n-1)-\Delta & \frac{\sqrt{2}}{2}g\sqrt{n} & 0 \\
\frac{\sqrt{2}}{2}g\sqrt{n} & \omega n & \frac{\sqrt{2}}{2}g\sqrt{n+1} \\
0 & \frac{\sqrt{2}}{2}g\sqrt{n+1} & \omega (n+1)+\Delta
\end{array}
\right) .  \nonumber \\
\end{eqnarray}
In the RWA, one can diagonalize the above Hamiltonian easily.

Including the counter rotating-wave terms, the total photonic number is not
conserved, the above subspace related to $n$ is not closed, rendering the
complication of the solution. Here, we present a treatment to the
Hamiltonian (~\ref{crtham}) based on the unitary transformation~\cite
{silbey,zheng,yu,zhang1}: $H^{\prime }=exp(U)Hexp(-U)$ with the following
displaced operator
\begin{equation}
U=\exp \left[ \frac g\omega J_z\left( a^{\dagger }-a\right) \right] .
\end{equation}
The transformed Hamiltonian is
\begin{eqnarray}
H^{\prime } &=&H_0+H_1+H_2,  \label{transham1} \\
H_0 &=&\omega a^{\dagger }a-g^2/\omega J_z^2, \\
H_1 &=&\Delta J_xG_0\left( a^{\dagger }a\right) +iJ_y\Delta F_1\left(
a^{\dagger }a\right) (a^{\dagger }-a), \\
H_2 &=&\Delta J_x\{\cosh \left[ \frac g\omega \left( a^{\dagger }-a\right)
\right] -G_0\left( a^{\dagger }a\right) \}  \nonumber \\
&&+iJ_y\Delta \{\sinh \left[ \frac g\omega \left( a^{\dagger }-a\right)
\right] -F_1\left( a^{\dagger }a\right) (a^{\dagger }-a)\},  \nonumber \\
&&
\end{eqnarray}
where $G_0(a^{\dagger }a)$ denotes zero excitation of photon in state $%
|n\rangle $
\begin{eqnarray}
\langle n|G_0(a^{\dagger }a)|n\rangle  &=&\langle n|\cosh \left[ \frac
g\omega \left( a^{\dagger }-a\right) \right] |n\rangle   \nonumber \\
&=&e^{-\frac{g^2}{2\omega ^2}}L_n(\frac{g^2}{\omega ^2}),
\end{eqnarray}
with the Laguerre polynomials $L_n^{m-n}(x)=\sum_{i=0}^{\min
\{m,n\}}(-1)^{n-i}\frac{m!x^{n-i}}{(m-i)!(n-i)!i!}$. Note that  $F_1\left(
a^{\dagger }a\right) (a^{\dagger }-a)$ plays a role of creating and
eliminating a single photon. Since $aF_1\left( a^{\dagger }a\right) $ only
has value in $\langle n\left| n+1\right\rangle $ and the term $F_1\left(
a^{\dagger }a\right) a^{\dagger }$ only has value in $\langle n+1\left|
n\right\rangle $, so we  have the following overlap
\begin{eqnarray}
\langle n+1|F_1\left( a^{\dagger }a\right) a^{\dagger }\left| n\right\rangle
&=&\langle n+1|\sinh \left[ \frac g\omega \left( a^{\dagger }-a\right)
\right] |n\rangle   \nonumber \\
&=&\frac 1{\sqrt{n+1}}\frac g\omega e^{-\frac{g^2}{2\omega ^2}}L_n^1(\frac{%
g^2}{\omega ^2}).
\end{eqnarray}
Since $\cosh \left[ \frac g\omega \left( a^{\dagger }-a\right) \right] $ and
$\sinh \left[ \frac g\omega \left( a^{\dagger }-a\right) \right] $ contain
powers of the number operator $a^{\dagger }a$ with even and odd functions
respectively, they appear in $H_2,\;$as  $\cosh \left[ \frac g\omega \left(
a^{\dagger }-a\right) \right] =G_0\left( a^{\dagger }a\right) +O(\frac{g^2}{%
\omega ^2})$ and $\sinh \left[ \frac g\omega \left( a^{\dagger }-a\right)
\right] =F_1\left( a^{\dagger }a\right) (a^{\dagger }-a)+O(\frac{g^3}{\omega
^3})$, where higher terms describing the double  and multi-photon transition
processes are neglected. Thus we have, $H^{\prime }=H_0+H_1$, similar to the
approximation performed in the single-qubit Rabi model~\cite{zhang1}.

\begin{center}
\begin{figure*}[tbp]
\includegraphics[width=\textwidth]{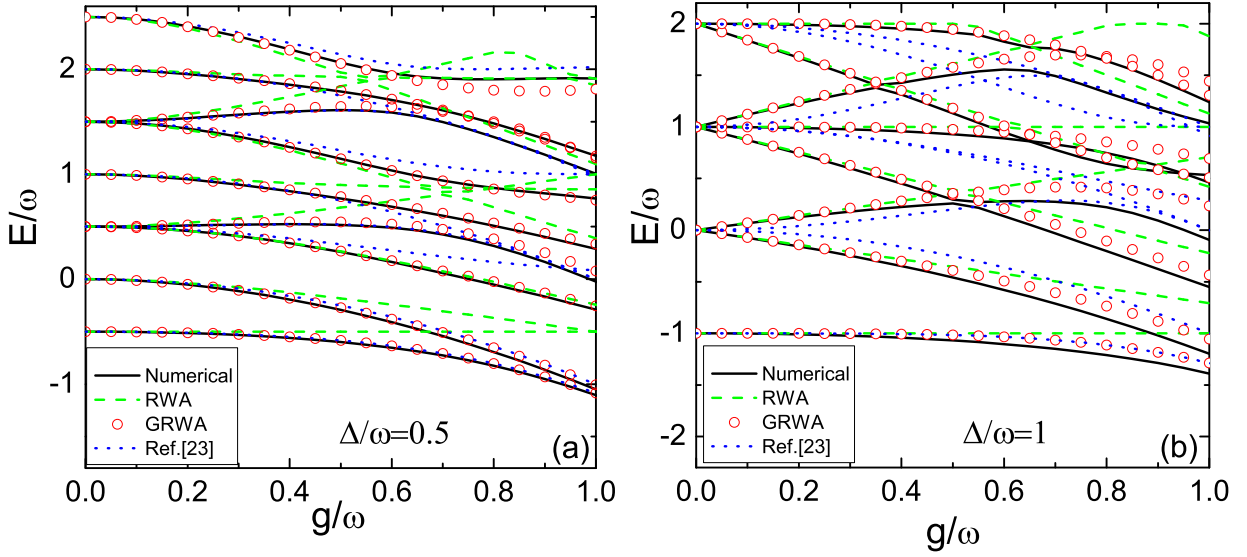}
\caption{We plot the ground state energy $E_{0}$ in Eq.(\ref{gs}), the first
excited state energy $E_{1\pm }$ in Eq.(\ref{1-2th}), and the energies
obtained by solving Eq.(\ref{GRWAmatrix}) for $n>0$ by the GRWA method
(circles) for different $\Delta /\omega =0.5$ (a), $\Delta /\omega =1$ (b).
And the energies obtained from numerical exact diagonalization (solid
lines), results of RWA in Eq.(~\ref{RWA}) and results in Ref.~\protect\cite
{agarwal} expressed in Eq.(\ref{zeroenergy})obtained by zero-order
approximation are plotted for comparison. }
\label{energy level}
\end{figure*}
\end{center}

As the zeroth-order approximation, we neglect the terms $F_1\left(
a^{\dagger }a\right) (a^{\dagger }-a)$ involving creating and eliminating a
single photon, the Hamiltonian is then approximated as
\begin{equation}
H^{^{\prime }}=\omega a^{\dagger }a-g^2/\omega J_z^2+\Delta J_xG_0\left(
a^{\dagger }a\right) .
\end{equation}
In the spin and photonic basis of $|1,n\rangle ,|0,n\rangle $ and $%
|-1,n\rangle $, $\ $we have
\[
H^{\prime }=\left(
\begin{array}{ccc}
\omega n-\frac{g^2}\omega  & \frac \Delta {\sqrt{2}}G_0(n) & 0 \\
\frac \Delta {\sqrt{2}}G_0(n) & \omega n & \frac \Delta {\sqrt{2}}G_0(n) \\
0 & \frac \Delta {\sqrt{2}}G_0(n) & \omega n-\frac{g^2}\omega
\end{array}
\right) .
\]
The corresponding eigenvalues and eigenfunctions are straightforwardly given
by
\begin{eqnarray}  \label{zeroenergy}
\varepsilon _{\pm ,n} &=&\frac \omega 2(2n-\frac{g^2}{\omega ^2}\pm \sqrt{%
(\frac g\omega )^4+4[\frac{\Delta G_0(n)}\omega ]^2},  \nonumber
\label{zeroenergy} \\
\varepsilon _{0,n} &=&\omega n-\frac{g^2}\omega ,
\end{eqnarray}
and
\begin{equation}
|\varepsilon _{0,n}\rangle =\left(
\begin{array}{c}
1 \\
0 \\
-1
\end{array}
\right) ,|\varepsilon _{\pm ,n}\rangle =\left(
\begin{array}{c}
1 \\
(\chi _n\pm \sqrt{8+\chi _n^2})/2 \\
1
\end{array}
\right) ,  \label{zerostates}
\end{equation}
where $\chi _n=\frac{\sqrt{2}g^2}{\omega \Delta G_0(n)}$. Interestingly, the
eigenvalues and eigenstates obtained in this way are exactly the same as
those obtained by the adiabatic approximation~\cite{agarwal}. The
zeroth-order energy spectrum is plotted in Fig.~\ref{energy level} with blue
dashed lines. For comparison, the energies obtained from numerical exact
diagonalization and in the RWA are also given with black solid lines and
green dashed lines. The ground-state energy and low excited energies agree
well with the numerical results for $\Delta /\omega =0.5$. It is obvious
that the RWA results become worse in the strong coupling regime. The
adiabatic approximate results also deviate from the numerical ones in the
ultrastrong coupling regime, and this situation becomes more serious with
increasing atomic transition frequency. Neglecting the term $iJ_yF_1\left(
a^{\dagger }a\right) (a^{\dagger }-a)$ in the zeroth order approximation,
there only exist transition between states with the same values of
oscillator excitation, $|0,n\rangle $ and $|\pm 1,n\rangle $. Hence, the
validity of the adiabatic approximation is restricted to the large detuning
regime $\Delta \ll \omega $. The transitions between various states with
different values of oscillator excitation for large value of $\Delta $ will
be considered in the next section.

\section{Generalized rotating-wave approximation}

As the first-order approximation, the term $iJ_yF_1\left( a^{\dagger
}a\right) (a^{\dagger }-a)$ will be included, so the Hamiltonian now
consists of two parts
\begin{eqnarray}
H_0^{^{\prime }} &=&\omega a^{\dagger }a-\frac{g^2}\omega J_z^2+\Delta \beta
J_x, \\
H_1^{^{\prime }} &=&\Delta J_x[G_0\left( a^{\dagger }a\right) -\beta ]+\frac
\Delta 2F_1\left( a^{\dagger }a\right) (a^{\dagger }-a)(J_{+}-J_{-})
\end{eqnarray}
where $\beta =G_0\left( 0\right) =e^{-\frac{g^2}{2\omega ^2}}$.

Obviously, the spin and photons in $H_0^{\prime }$ are decoupled and its
spin part can be diagonalized in the spin basis of $|-1\rangle ,|0\rangle $
and $|1\rangle $ by a unitary matrix $S$ as
\begin{equation}
S=\left(
\begin{array}{ccc}
1/\lambda _{-} & 1/\sqrt{2} & 1/\lambda _{+} \\
\mu _{-}/\lambda _{-} & 0 & \mu _{+}/\lambda _{+} \\
1/\lambda _{-} & -1/\sqrt{2} & 1/\lambda _{+}
\end{array}
\right) ,
\end{equation}
where $\mu _{\pm }=\frac{\chi _0}2\pm \frac{\sqrt{\chi _0^2+8}}2,\chi _0=%
\frac{\sqrt{2}g^2}{\omega \Delta \beta },\lambda _{\pm }=\sqrt{2+\mu _{\pm
}^2}$. The corresponding eigenvalues are $\varepsilon _{\pm }=\frac{\Delta
\beta }{2\sqrt{2}}(-\chi _0\pm \sqrt{\chi _0^2+8})$ and $\varepsilon _0=-%
\frac{g^2}\omega $. Therefore the diagonal $H_0^{\prime }$ takes the form
\begin{equation}
\stackrel{\sim }{H_0}=\left(
\begin{array}{ccc}
\omega n+\varepsilon _{-} & 0 & 0 \\
0 & \omega n+\varepsilon _0 & 0 \\
0 & 0 & \omega n+\varepsilon _{+}
\end{array}
\right) ,
\end{equation}
The first order term $H_1^{^{\prime }}$ is transformed by the unitary matrix
\begin{eqnarray}
\stackrel{\sim }{H_1} &=&S^{+}H_1^{\prime }S  \nonumber \\
&=&\left(
\begin{array}{ccc}
\frac{2\sqrt{2}\mu _{-}}{\lambda _{-}^2} & 0 & \frac{\sqrt{2}(\mu _{+}+\mu
_{-})}{\lambda _{+}\lambda _{-}} \\
0 & 0 & 0 \\
\frac{\sqrt{2}(\mu _{+}+\mu _{-})}{\lambda _{+}\lambda _{-}} & 0 & \frac{2%
\sqrt{2}\mu _{+}}{\lambda _{+}^2}
\end{array}
\right) \Delta [G_0\left( a^{\dagger }a\right) -\beta ]  \nonumber \\
&&+\left(
\begin{array}{ccc}
0 & -\frac{\mu _{-}}{\lambda _{-}} & 0 \\
\frac{\mu _{-}}{\lambda _{-}} & 0 & \frac{\mu _{+}}{\lambda _{+}} \\
0 & -\frac{\mu _{+}}{\lambda _{+}} & 0
\end{array}
\right) \Delta F_1\left( a^{\dagger }a\right) (a^{\dagger }-a).
\end{eqnarray}
Neglecting the counter rotating-wave terms $a^{\dagger }J_{+}+aJ_{-}$ and
the remote matrix elements $\frac{\sqrt{2}(\mu _{+}+\mu _{-})}{\lambda
_{+}\lambda _{-}}$, we give  the total  Hamiltonian as
\begin{eqnarray}
H_{\mathtt{GRWA}} &=&\omega a^{\dagger }a+\{\varepsilon _{+}+\frac{2\sqrt{2}%
\mu _{+}\Delta }{\lambda _{+}^2}[G_0\left( a^{\dagger }a\right) -\beta %
]\}|1\rangle \langle 1|  \nonumber  \label{GRWAham} \\
&&+\{\varepsilon _{-}+\frac{2\sqrt{2}\mu _{-}\Delta }{\lambda _{-}^2}[%
G_0\left( a^{\dagger }a\right) -\beta ]\}|-1\rangle \langle -1|  \nonumber \\
&&+\varepsilon _0|0\rangle \langle 0|+\frac{\mu _{+}\Delta }{\lambda _{+}}%
F_1\left( a^{\dagger }a\right) (a|1\rangle \langle 0|+a^{\dagger }|0\rangle
\langle 1|)  \nonumber \\
&&-\frac{\mu _{-}\Delta }{\lambda _{-}}F_1\left( a^{\dagger }a\right)
(a|0\rangle \langle -1|+a^{\dagger }|-1\rangle \langle 0|),
\end{eqnarray}
where there is only the energy-conserving term is $a|1\rangle \langle 0|+h.c$%
, $a^{\dagger }|-1\rangle \langle 0|+h.c$ with  renormalized coefficients  $%
\frac{\mu _{+}\Delta }{\lambda _{+}}F_1\left( a^{\dagger }a\right) $and $-%
\frac{\mu _{-}\Delta }{\lambda _{-}}F_1\left( a^{\dagger }a\right) $. So it
is   exactly same as the Tavis-Cummings model with a renormalized parameters in the
RWA form. In this sense, we can also call the first-order approximation as
GRWA.  The effect of the  counterrotating-wave interaction in the original
model, which play a role in the ultrastrong coupling regime,  now is
absorbed in  $iJ_yF_1\left( a^{\dagger }a\right) (a^{\dagger }-a)$.

In the basis of $|-1,n+1\rangle $, $|0,n\rangle $ and$|1,n-1\rangle
,(n=1,2,...),$ $H_{\mathtt{GRWA}}$ takes the following matrix form
\begin{widetext}
\begin{equation}\label{GRWAmatrix}
H_{\texttt{GRWA}}=\left(
\begin{array}{ccc}
\omega (n+1)+\xi _{-,n+1} & -\frac{\mu _{-} }{\lambda _{-}}R_{n,n+1}%
\sqrt{n+1} & 0 \\
-\frac{\mu _{-} }{\lambda _{-}}R_{n,n+1}\sqrt{n+1} & \omega
n+\varepsilon _{0} & \frac{\mu _{+} }{\lambda _{+}}R_{n-1,n}\sqrt{n}
\\
0 & \frac{\mu _{+}}{\lambda _{+}}R_{n-1,n}\sqrt{n} & \omega (n-1)+\xi
_{+,n-1}%
\end{array}%
\right) ,
\end{equation}
\end{widetext}
where $\xi _{+,n-1}=\varepsilon _{+}+\frac{2\sqrt{2}\mu _{+}\Delta
[G_0\left( n-1\right) -\beta ]}{\lambda _{+}^2}$, $\xi _{-,n+1}=\varepsilon
_{-}+\frac{2\sqrt{2}\mu _{-}\Delta [G_0\left( n+1\right) -\beta ]}{\lambda
_{-}^2}$ , and $R_{n,n+1}=\Delta \langle n|F_1\left( a^{\dagger }a\right)
a|n+1\rangle $, $R_{n-1,n}=\Delta \langle n-1|F_1\left( a^{\dagger }a\right)
a|n\rangle $. Similar to the usual RWA Hamiltonian(~\ref{RWA}), the
eigenstates $|\phi _n\rangle $ and eigenvalues $E_n$ of the GRWA one can be
easily obtained.

For $n=0$,  in the basis $|-1,1\rangle $ and $|0,0\rangle ,$ we have
\begin{equation}
H_{\mathtt{GRWA}}=\left(
\begin{array}{cc}
\varepsilon _0 & -\frac{\mu _{-}R_{0,1}}{\lambda _{-}} \\
-\frac{\mu _{-}R_{0,1}}{\lambda _{-}} & \omega +\xi _{-,1}
\end{array}
\right) .
\end{equation}
which results in  the first- and second-excited eigenvalues
\begin{eqnarray}
E_{1,\pm } &=&\frac{\varepsilon _0+\omega +\xi _{-}}2  \label{1-2th} \\
&&\pm \frac 12\sqrt{(\varepsilon _0-\omega -\xi _{-})^2+4(\frac{\mu
_{-}\Delta R_{0,1}}{\lambda _{-}})^2}.
\end{eqnarray}
and eigenstates $|\phi \rangle _{\pm }=\{\frac{\lambda _{-}}{2\mu _{-}\Delta
R_{0,1}}[(\varepsilon _0-\omega -\xi _{-})\pm \sqrt{(\varepsilon _0-\omega
-\xi _{-})^2+4(\frac{\mu _{-}\Delta R_{0,1}}{\lambda _{-}})^2}%
]\}|-1,1\rangle +|0,0\rangle $.

The ground state energy for the state $|-1,0\rangle $ is
\begin{equation}
E_0=\frac{\Delta \beta }{2\sqrt{2}}(-\chi \pm \sqrt{\chi ^2+8}).  \label{gs}
\end{equation}

\begin{center}
\begin{figure*}[tbp]
\includegraphics[scale=1]{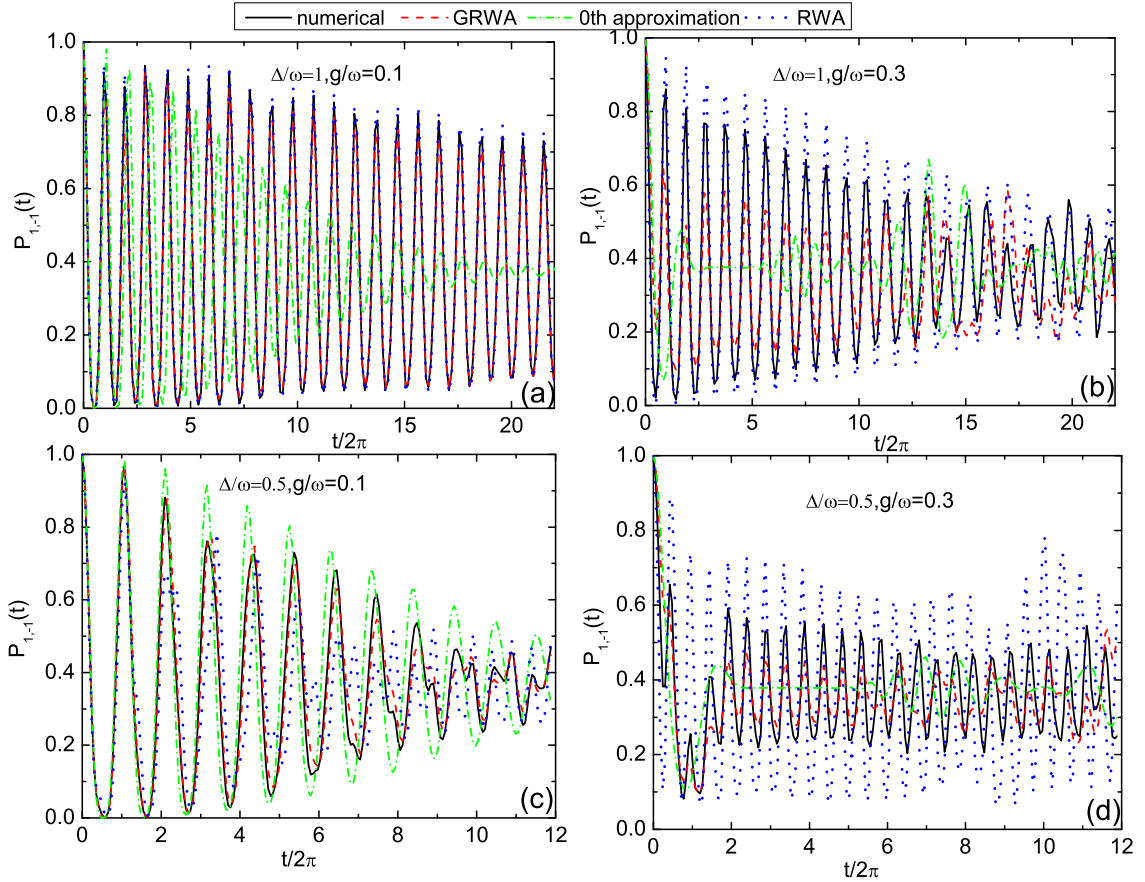}
\caption{Population dynamics for $P_{1,-1}(t)$ of GRWA, given $%
\Delta/\omega=1$, $g/\omega=0.1$ (a), $\Delta/\omega=1$, $g/\omega=0.3$ (b),
$\Delta/\omega=0.5$, $g/\omega=0.1$ (c), and $\Delta/\omega=0.5$, $g/\omega=1
$ (d). We choose $n=\langle\alpha\rangle=4$. For comparison, we plot results
obtained by numerical exact diagonalization (solid lines), that of the RWA
in Eq.(~\ref{RWA}) (dots), and by the zero-order approximation in Eq.(\ref
{zeroenergy}).}
\label{inversion dynamics}
\end{figure*}
\end{center}

Figure.~\ref{energy level} shows  the  energy level $E/\omega $ as a
function of the coupling strength $g/\omega $ for $\Delta /\omega =0.5$ and $%
\Delta /\omega =1$ within different approaches.  It is obvious that the
GRWA results for the energy is much better than but the adiabatic
approximation one~\cite{agarwal}, comparing with that in the numerical
exact diagonalization.  Remarkably, the GRWA works reasonably well even for
large detunings with $\Delta /\omega =0.5$. As illustrated in Fig.~\ref
{energy level}(a), the ground state energy $E_0$ in Eq.(\ref{gs}) agrees
well with the numerical results in the whole coupling regime and there is
qualitative agreement for high  energy levels. The RWA reproduces the
correct limiting behavior as $g/\omega \rightarrow 0$, but breaks down in
the strong coupling regime  $g/\omega \geq 0.3$. The RWA requires small
detuning and the adiabatic approximation in Ref.~\cite{agarwal} is derived
under the assumption that $\Delta \ll \omega ,$and the effect of the counter
rotating-wave terms is totally ignored. Our approach is basically a
perturbation in $\Delta /\omega ,$ the adiabatic approximated one is
actually the zero-order perturbation with the framework of the present
approach, the GRWA is the first-order perturbation one, so as the increase
of $\Delta /\omega $, the present GRWA becomes better. In both  the GRWA and
the adiabatic approximation, the  effect of the counter rotating-wave terms
is partially included.

\section{Population dynamics}

\bigskip The collapse and revival behavior for a single-qubit case was
studied ~\cite{Braak,zhang1,grifoni} and we explore the atomic population
inversion in the two-qubit cavity system. Here we apply the eigenvalues and
eigenstates obtained by GRWA to investigate the problem in all coupling
regimes. To study the population dynamics, we need  the eigenstates for the
original Hamiltonian (1) with counter rotating-wave terms, which can be
obtained using a unitary transformation in zero order approximation as
\begin{eqnarray}
|\varphi _{0,n}^0\rangle  &=&U^{\dagger }|\varepsilon _{0,n}\rangle =\left(
\begin{array}{c}
|n\rangle _1 \\
0 \\
-|n\rangle _{-1}
\end{array}
\right) ,  \nonumber  \label{zerostate1} \\
|\varphi _{\pm ,n}^0\rangle  &=&U^{\dagger }|\varepsilon _{\pm ,n}\rangle
=\left(
\begin{array}{c}
|n\rangle _1 \\
(\chi \pm \sqrt{8+\chi ^2})/2|n\rangle _0 \\
|n\rangle _{-1}
\end{array}
\right) ,
\end{eqnarray}
where the oscillator states $|n\rangle _j=\exp [\frac{jg}\omega (a^{\dagger
}-a)]|n\rangle ,j=0,\pm 1$ are called extended coherent states. Similarly,
under the first-order approximation the original eigenstates are evaluated
as $|\varphi _n^1\rangle =U^{\dagger }S^{\dagger }|\phi _n\rangle $.

The initial state is set $|\varphi (0)\rangle =|-1\rangle |\alpha
_{-1}\rangle $ with $|\alpha _{-1}\rangle =e^{g/\omega (a^{\dagger
}-a)}|\alpha \rangle $. The wavefuntion evolutes as $|\varphi (t)\rangle
=e^{-iHt}|\varphi (0)\rangle $, which can be expanded by the eigenvalues and
eigenstates for the original Hamiltonian under the zeroth- and first-order
approximation.

The population for the qubits remain in the state $|1,-1\rangle $ is
expressed as
\begin{equation}
P_{1,-1}(t)=|\langle -1|\mathtt{Tr}_{\mathtt{ph}}|\varphi (t)\rangle \langle
\varphi (t)|-1\rangle |^2.
\end{equation}
This expectation value with zeroth-order approximation and the GRWA method
are plotted respectively in Fig.~\ref{inversion dynamics} for coupling
strength $g/\omega =0.1$, $0.3$ with different $\Delta /\omega =1$ and $0.5$%
. For comparison, the results from the RWA and numerical exact
diagonalization are also collected. Obviously, the population inversion
results of the GRWA agree well with the numerical ones. And there is
substantial improvements over those obtained by the zeroth order
approximation in the ultrastrong coupling regime. It is ascribe to the
counterrotating-wave interaction in the first order correction, including
the states transition with different oscillator excitations, demonstrating
the validity of the eigenstates and eigenvalues in the ultrastrong coupling
regime by the GRWA.

\section{conclusion}

In summary,  the effective solvable Hamiltonian for the two-qubit
Tavis-Cummings model beyond RWA is derived by a unitary transformation,
which can in turn gives accurate eigenvalues and eigenstates. The zeroth-
order approximation  produce the analytical eigenvalues and eigenstates of
the adiabatic approximation completely. The first-order approximation,
called GRWA are mainly performed, where the rotating-wave interacting
coupling strength is renormalized and a counter rotating-wave interactions
are including the renormalized coefficients.  In the GRWA, the mathematical
simplicity of the ordinary RWA is retained, which facilitate the further
study.  The obtained energy spectrum  are in good agreement with the
numerical exact diagonalization ones in a wide range of coupling strength,
much better than  the previous  adiabatic approximation. The population
inversion obtained using GRWA is also quantitative agreement with the
numerical ones, indicating  the valid  eigenstates and eigenvalues in the
ultrastrong coupling regime for different detuning regime. By the analytical
eigensolutions, all properties for this two-qubit cavity coupling system can
be easily explored. Our approach can be extended to the multi-qubit case,
such as the  Dicke model.

\acknowledgments
This work was supported by National Natural Science Foundation of China
(Grants No.~11174254 and No.~11104363), and Research Fund for the
Central Universities(No. CQDXWL-2013-Z014 and No. CQDXWL-2012-Z005).

%$^{*}$ Corresponding author. Email:yuyuzh@cqu.edu.cn.

\end{document}